\documentclass[%
 reprint,
 superscriptaddress,
 amsmath,amssymb,
 aps,
]{revtex4-2}
\usepackage{graphicx}
\usepackage[margin=01in]{geometry}
\usepackage{multirow}
\usepackage{csquotes}

\begin{document}
\title{Ductility mechanisms in complex concentrated refractory alloys from atomistic fracture simulations}

\author{Wenqing Wang}
\affiliation{Materials Sciences Division, Lawrence Berkeley National Laboratory, Berkeley, California 94720, USA}
\affiliation{Department of Materials Science and Engineering, University of California, Berkeley, California 94720, USA}

\author{Punit Kumar}
\affiliation{Materials Sciences Division, Lawrence Berkeley National Laboratory, Berkeley, California 94720, USA}
\affiliation{Department of Materials Science and Engineering, University of California, Berkeley, California 94720, USA}

\author{David H. Cook}
\affiliation{Materials Sciences Division, Lawrence Berkeley National Laboratory, Berkeley, California 94720, USA}
\affiliation{Department of Materials Science and Engineering, University of California, Berkeley, California 94720, USA}

\author{Flynn Walsh}
\affiliation{Materials Sciences Division, Lawrence Berkeley National Laboratory, Berkeley, California 94720, USA}

\author{Buyu Zhang}
\affiliation{Materials Sciences Division, Lawrence Berkeley National Laboratory, Berkeley, California 94720, USA}
\affiliation{Department of Materials Science and Engineering, University of California, Berkeley, California 94720, USA}

\author{Pedro P.P.O. Borges}
\affiliation{Materials Sciences Division, Lawrence Berkeley National Laboratory, Berkeley, California 94720, USA}
\affiliation{Department of Materials Science and Engineering, University of California, Berkeley, California 94720, USA}

\author{Diana Farkas}
\affiliation{Department of Materials Science and Engineering, Virginia Polytechnic Institute, Computer Simulation Laboratory, 201 Holden Hall, Blacksburg, VA 24061-0237, USA}

\author{Robert O. Ritchie} 
\email{roritchie@lbl.gov}
\affiliation{Materials Sciences Division, Lawrence Berkeley National Laboratory, Berkeley, California 94720, USA}
\affiliation{Department of Materials Science and Engineering, University of California, Berkeley, California 94720, USA}

\author{Mark Asta}
\email{mdasta@berkeley.edu}
\affiliation{Materials Sciences Division, Lawrence Berkeley National Laboratory, Berkeley, California 94720, USA}
\affiliation{Department of Materials Science and Engineering, University of California, Berkeley, California 94720, USA}

\begin{abstract} 
The striking variation in damage tolerance among refractory complex concentrated alloys is examined through the analysis of atomistic fracture simulations, contrasting behavior in elemental
Nb with that in brittle NbMoTaW and ductile Nb$_{45}$Ta$_{25}$Ti$_{15}$Hf$_{15}$.  We employ machine-learning interatomic potentials (MLIPs), including a new MLIP 
developed for NbTaTiHf, in atomistic simulations of crack tip extension mechanisms based on analyses of atomistic fracture 
resistance curves. While the initial behavior of sharp cracks shows good correspondence with the Rice theory, fracture resistance curves 
reveal marked changes in fracture modes for the complex alloys as crack extension proceeds. In NbMoTaW, compositional complexity appears to promote dislocation nucleation relative to pure Nb, despite theoretical predictions that the alloy should be relatively more brittle. In Nb$_{45}$Ta$_{25}$Ti$_{15}$Hf$_{15}$, alloying not only changes the fracture mode relative to elemental Nb, but promotes dislocation accumulation at the crack tip, leading to higher resistance to crack propagation.
\end{abstract}

\maketitle

\section{Introduction}
Body-centered cubic (bcc) complex concentrated alloys containing refractory elements (hereafter referred to as refractory high entropy alloys [RHEAs]) have received great attention due to 
the retention of high yield strength at elevated temperatures displayed by several of these materials \cite{senkov2011mechanical,SENKOV20101758}. However, a significant number of 
of RHEAs, such as the well-established Nb-Mo-Ta-W based alloys, display inadequate tensile ductility and/or fracture toughness at lower temperatures 
\cite{han2018microstructures,kumar2023compressive, Punit2024}, imposing severe restrictions on their potential applications as structural materials. In contrast, Nb-Ta-Ti-Hf based 
RHEAs are found to possess an excellent combination of tensile ductility and high fracture toughness at ambient temperature \cite{fan2022remarkably,cook2024kink}. These differences in
behavior across RHEAs are striking given the chemical similarity of many of the alloys, e.g., brittle NbMoTaW and ductile Nb$_{45}$Ta$_{25}$Ti$_{15}$Hf$_{15}$Hf have valence electron
counts differing by only 0.8 electrons per atom.  In the design of RHEAs for structural applications, a central challenge remains understanding the mechanisms underlying the 
large variations in ductility arising from such subtle compositional changes.

Computational approaches have been used widely to investigate intrinsic ductility in RHEAs, through consideration of theoretical models analyzing the behavior of initially sharp crack tips 
under tensile loading. Specifically, the Rice-Thomson model categorizes the intrinsic ductility by analyzing whether an atomically sharp crack propagates by bond breaking (intrinsically 
brittle) or dislocation emission (intrinsically ductile) \cite{lin1986cleavage,rice1974ductile}. Rice further highlights the ratio of surface energy to the unstable stacking fault (USF)
energy \cite{rice1992dislocation} as a key parameter for intrinsic ductility \cite{rice1974ductile}. These models have been the basis for first-principles calculations in RHEA systems \cite{borges2024electronic,tandocmining} that have qualitatively reproduced trends in intrinsic ductility across compositions.  However, these investigations have not directly considered crack behavior beyond idealized 
initial conditions implicit in the Rice-Thomson analyses. Further understanding of intrinsic ductility and toughness requires knowledge of the crack tip behavior and geometry after 
initiation, and how these features reflect the complex interplay of dislocation slip and twinning deformation mechanisms versus simple bond breaking.

This study investigates such mechanisms using molecular statics atomistic simulations based on machine-learning interatomic potentials (MLIPs), which enable the simulation of deformation 
and fracture processes in complex concentrated alloys with near density-functional-theory (DFT) accuracy.  Similar atomistic simulations, often with more idealized potential models, have 
been applied extensively in studies of crack tip behavior in related alloys \cite{de2017ideal, qi2014tuning, andric2018atomistic,li2020unveiling,farkas2024role,cheung1994molecular}, 
yielding important insights into the origins of brittle-ductile transformations, crack-tip blunting by dislocation emission, twinning and nucleation of new grains 
\cite{farkas1998atomistic, farkas2005twinning, guo2007atomistic}. The current work extends such prior studies by employing MLIP models in analyses of atomistic fracture resistance ($G_R$)
curves and crack tip opening displacement (CTOD) 
as a function of crack extension. To gain insights into the role of
compositional complexity, and compositional variations across RHEAs, analyses are presented comparing elemental Nb, $\text{Nb}_{45}\text{Ta}_{25}\text{Ti}_{15}\text{Hf}_{15}$ and NbMoTaW.
The results of these analyses highlight how ductile fracture is enhanced by compositional complexity.  They further demonstrate that the crack propagation mode in RHEAs may change during 
the fracture process, with such changes being well captured by slope changes in the atomistic $G_R$ curve.


\section{Results}
\subsection{The Rice Theory Analysis}

We begin by considering differences between elemental Nb, MoNbTaW, and $\text{Nb}_{45}\text{Ta}_{25}\text{Ti}_{15}\text{Hf}_{15}$ systems based on the Rice theory \cite{rice1992dislocation}, 
which predicts a crossover condition, under pure mode I loading, for cleavage bond breaking (brittle behavior) and dislocation emission (ductile behavior) at an advancing crack tip. 
Specifically intrinsically brittle (ductile) behavior is predicted if the ratio ($\gamma_\text{\text{USF}} / \gamma_{\text{surface}}$) of USF energy
($\gamma_\text{\text{USF}}$) to surface energy ($\gamma_{\text{surface}}$) is larger (smaller) than a threshold value defined as:
\begin{equation}
\label{eq:Rice}
 \left[ \frac{\gamma_\text{\text{USF}}}{\gamma_{\text{surface}}} \right]_{\textrm{threshold}}=\frac{(1+\cos\theta)\sin^2\theta}{4(1+(1-\nu)\tan^2\phi)}
\end{equation}
where $\nu$ is the Poisson's ratio, $\theta$ is the inclination angle between the fracture plane and the dislocation glide plane, and $\phi$ is the inclination angle between the slip 
direction with the normal to the crack tip. 

\begin{table*}
\caption{Values of the Rice ratio, $\gamma_\text{\text{USF}} / \gamma_{\text{surface}}$, and the approximated threshold values (Eq. \ref{eq:Rice}) for ductile/brittle fracture for elemental Nb, and the RHEAs $\text{Nb}_{45}\text{Ta}_{25}\text{Ti}_{15}\text{Hf}_{15}$ and NbMoTaW.  Results are given for $\{110\}$ and $\{112\}$ fracture planes, calculated using the MLIP of Yin et al. \cite{yin2021atomistic} for Nb and NbMoTaW and a new MLIP developed in this work for $\text{Nb}_{45}\text{Ta}_{25}\text{Ti}_{15}\text{Hf}_{15}$).}
\begin{ruledtabular}
\begin{tabular}{c|ccc|c}
Fracture & & $\gamma_\text{\text{USF}} / \gamma_{\text{surface}}$ &  & $\left[ \gamma_\text{\text{USF}} / \gamma_{\text{surface}} \right]_{\textrm{threshold}}$ \\
Plane & Nb & NbMoTaW & $\text{Nb}_{45}\text{Ta}_{25}\text{Ti}_{15}\text{Hf}_{15}$ & \\
\hline
$\{110\}$ & 0.391 & 0.526 & 0.299 & 0.263 \\
$\{112\}$ & 0.416 & 0.547 & 0.294 & 0.296

\label{tab:R-T}
\end{tabular}
\end{ruledtabular}
\end{table*}

For both $\{110\}$ and $\{112\}$ fracture planes, Table I presents a comparison of values for $\gamma_\text{\text{USF}}/{\gamma_{\text{surface}}}$, and the estimated threshold values 
$\left[\gamma_\text{\text{USF}}/{\gamma_{\text{surface}}}\right]_{\textrm{threshold}}$ calculated from the MLIP due to Yin et al. \cite{yin2021atomistic} for NbMoTaW and Nb, and the MLIP developed here
for $\text{Nb}_{45}\text{Ta}_{25}\text{Ti}_{15}\text{Hf}_{15}$ (see Methods section). In calculating these values we derive $\nu$ from the calculated single-crystal elastic constants
using Voigt-Reuss-Hill (VRH) averaging, and consider all possible $\{110\}$ and $\{112\}$ slip planes (see Table SIV), corresponding to those with a value of $\theta$ 
and $\phi$ that maximizes $\left[\gamma_\text{\text{USF}}/{\gamma_{\text{surface}}}\right]_{\textrm{threshold}}$.  Further, calculated $\gamma_\text{\text{USF}}$ results represent average
values that do not account for local variations due to compositional fluctuations in the RHEAs. Due to these approximations we do not expect the properties in Table I to give a rigorous
threshold condition for brittle versus ductile behavior, but the relative values provide a guide to what is expected in the atomistic fracture simulations presented in the next 
sub-section. Specifically, the Rice criterion with the values in Table I predicts brittle cleavage bond breaking fracture behavior for NbMoTaW and Nb.  
For $\text{Nb}_{45}\text{Ta}_{25}\text{Ti}_{15}\text{Hf}_{15}$, $\gamma_\text{\text{USF}}/{\gamma_{\text{surface}}}$ are close to threshold values, and ductile dislocation 
may be expected.  Additionally, the numbers in Table I suggest that the effect of adding group VI elements (Mo 
and W) to Nb is to raise the tendency toward brittle fracture, while the addition of group IV elements (Ti and Hf) is to increase ductility, consistent with trends reported previously
from first-principles calculations \cite{borges2024electronic, rao2019modeling, tsuru2024intrinsic}.

\subsection{Atomistic Simulations of Crack Propagation}  

We consider next direct atomistic simulations of crack-tip propagation using the $K$-test geometry with a semi-infinite crack \cite{andric2018atomistic} (see Methods section).
Simulation results are presented for four crack-tip orientations, corresponding to three crack line directions ($\langle1\bar{1}0\rangle$, $\langle1\bar{1}1\rangle$, and
$\langle1\bar{1}2\rangle$) with a $\{110\}$ fracture plane, and one crack line direction ($\langle1\bar{1}0\rangle$) for a $\{112\}$ fracture plane.

\begin{figure*}
    \centering
    \includegraphics[width = 1.0\textwidth]{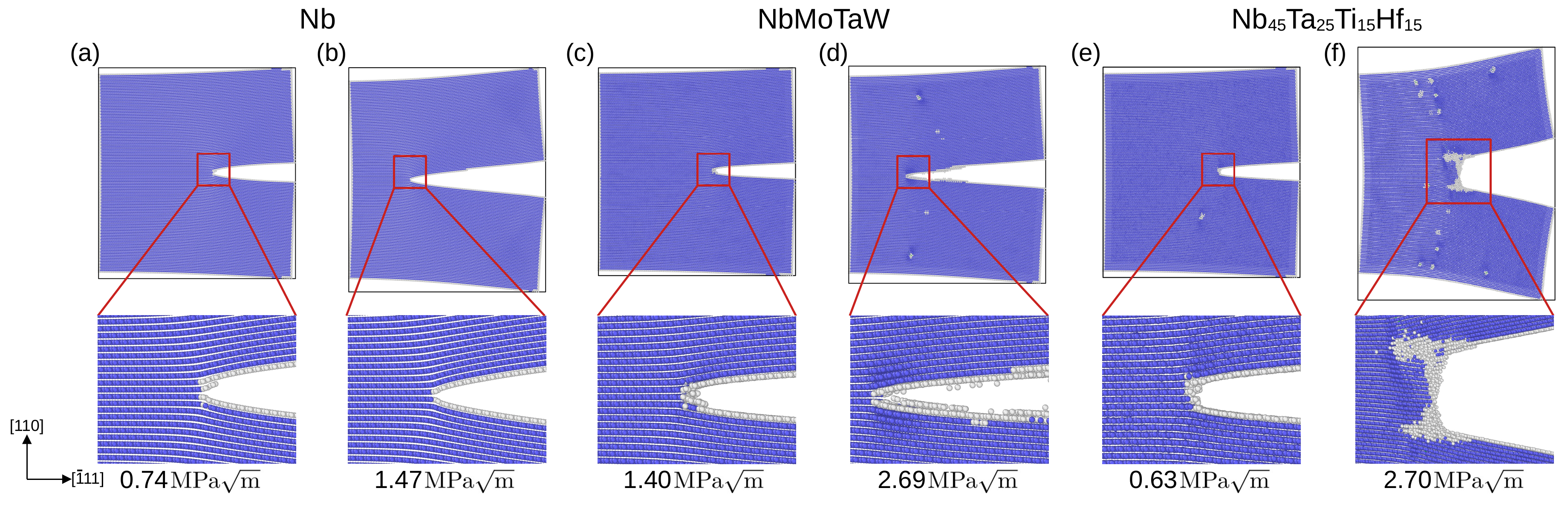}
    \caption{For a $\{110\}\langle1\bar{1}2\rangle$ crack orientation, atomic configurations from fracture simulations for Nb (a,b), NbMoTaW (c,d) and $\text{Nb}_{45}\text{Ta}_{25}\text{Ti}_{15}\text{Hf}_{15}$ (e,f), at different loadings represented by the values of $K_I$ listed below each panel.  Crack initiation events for an atomically sharp crack tip are provided in (a,c,e) and for later stages of crack propagation (b,d,f). The crack initiates by bond breaking in Nb and NbMoTaW, and by dislocation emission in $\text{Nb}_{45}\text{Ta}_{25}\text{Ti}_{15}\text{Hf}_{15}$. The crack initiates by bond breaking in Nb and NbMoTaW, and dislocation emissions in $\text{Nb}_{45}\text{Ta}_{25}\text{Ti}_{15}\text{Hf}_{15}$. While the fracture mode remained the same in Nb and $\text{Nb}_{45}\text{Ta}_{25}\text{Ti}_{15}\text{Hf}_{15}$, it displays dislocation emissions during some of the later stages in NbMoTaW. The atoms are colored by their structural types identified by the DXA algorithm\cite{stukowski2012automated} in Ovito \cite{stukowski2009visualization}: blue represents atoms in bcc environments and white represents atoms in defected environments.}
    \label{fig:110_121}
\end{figure*}

Figure \ref{fig:110_121} contrasts simulation results for loading of $\{110\}\langle1\bar{1}2\rangle$ oriented cracks for pure Nb ((a) and (b), and the RHEAs MoNbTaW ((b) and (c)) and
$\text{Nb}_{45}\text{Ta}_{25}\text{Ti}_{15}\text{Hf}_{15}$ ((d) and (e)). For pure Nb, the crack is observed to propagate by bond breaking, indicating brittle fracture behavior, 
consistent with the Rice theory predictions from Table I.  For NbMoTaW crack propagation for the first 50 $\AA$ proceeds through a mixed mode of bond breaking and 
dislocation emissions, but switches to bond breaking at later stages. Thus, despite the results in Table I, indicating more brittle behavior for NbMoTaW relative to Nb, plasticity is 
observed in the initial stages of crack propagation for the former.  This behavior may represent an effect of fluctuations associated with compositional complexity, giving rise to 
local environments with lower barriers for dislocation nucleation than would be expected from the average $\gamma_\text{\text{USF}}$ value.

Turning to the results for $\text{Nb}_{45}\text{Ta}_{25}\text{Ti}_{15}\text{Hf}_{15}$ in Fig. \ref{fig:110_121} (d) and (e), the initial crack propagation is associated with a 
dislocation emission event, and further crack advances lead to more dislocation emissions, significant blunting of the crack tip and twin formation.  These observations reflect 
a ductile fracture mode, correlating with the near threshold values for $\gamma_\text{\text{USF}}/{\gamma_{\text{surface}}}$ in Table I.  For the remaining crack tip orientations we
focus on comparisons of results for MoNbTaW and $\text{Nb}_{45}\text{Ta}_{25}\text{Ti}_{15}\text{Hf}_{15}$.  

\begin{figure*}
    \centering
    \includegraphics[width = 0.9\textwidth]{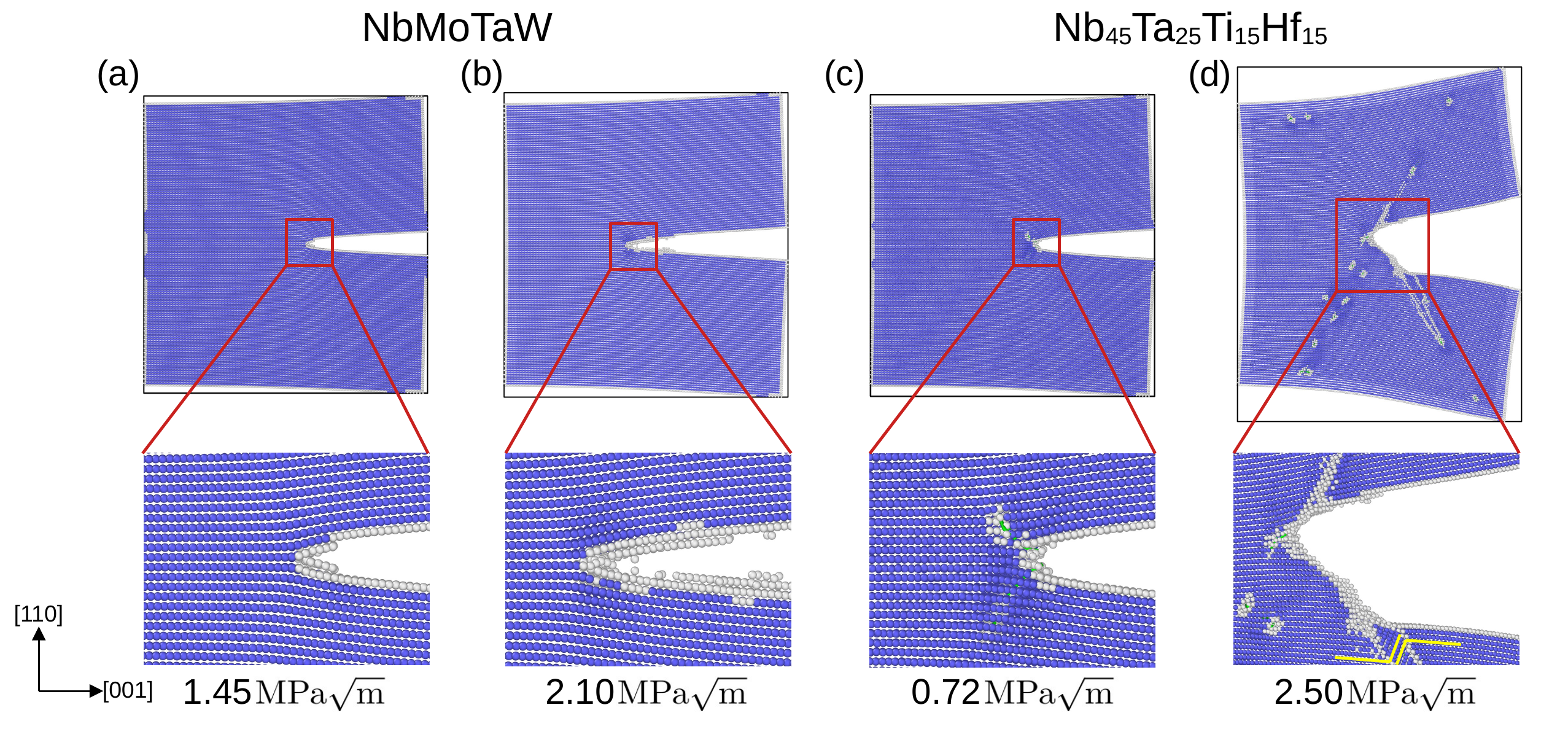}
    \caption{For a $\{110\}\langle1\bar{1}0\rangle$ crack orientation, atomic configurations from fracture simulations for NbMoTaW (a,b) and $\text{Nb}_{45}\text{Ta}_{25}\text{Ti}_{15}\text{Hf}_{15}$ (c,d), at different loadings represented by the values of $K_I$ listed below each panel.  Crack initiation events for an atomically sharp crack tip are provided in (a,c) and for later stages of crack propagation (b,d). The crack initiates by bond breaking in NbMoTaW and by dislocation emission in $\text{Nb}_{45}\text{Ta}_{25}\text{Ti}_{15}\text{Hf}_{15}$. The fracture modes remained the same for both RHEAs in later stages of crack propagation. The atoms are colored in the same manner as in Fig. \ref{fig:110_121}.  The golden lines at the bottom of panel (d) mark atomic planes to illustrate the formation of a deformation twin.}
    \label{fig:110_110}
\end{figure*}

\begin{figure*}
    \centering
    \includegraphics[width = 0.9\textwidth]{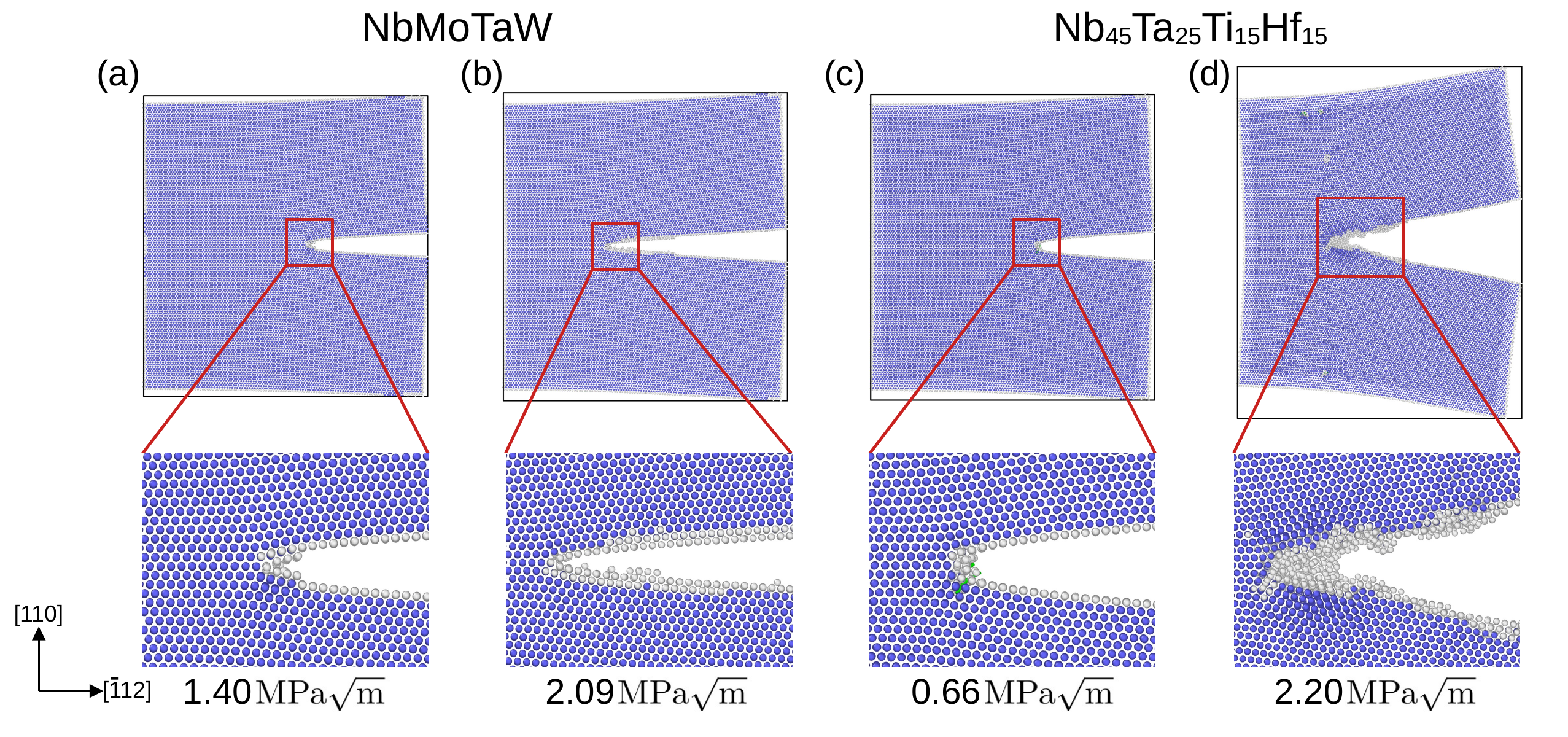}
    \caption{For a $\{110\}\langle1\bar{1}1\rangle$ crack orientation, atomic configurations from fracture simulations for NbMoTaW (a,b) and $\text{Nb}_{45}\text{Ta}_{25}\text{Ti}_{15}\text{Hf}_{15}$ (c,d), at different loadings represented by the values of $K_I$ listed below each panel.  Crack initiation events for an atomically sharp crack tip are provided in (a,c) and for later stages of crack propagation (b,d). The crack initiates by bond breaking in NbMoTaW and by dislocation emissions in $\text{Nb}_{45}\text{Ta}_{25}\text{Ti}_{15}\text{Hf}_{15}$. While the fracture mode in NbMoTaW remained the same in the later stages, the fracture mode in $\text{Nb}_{45}\text{Ta}_{25}\text{Ti}_{15}\text{Hf}_{15}$ switches to predominantly bond breaking.  The atoms are colored in the same manner as in Fig. \ref{fig:110_121}.}
    \label{fig:110_111}
\end{figure*}

Figures \ref{fig:110_110} and \ref{fig:110_111} show representative atomistic configurations for crack propagation in NbMoTaW and 
$\text{Nb}_{45}\text{Ta}_{25}\text{Ti}_{15}\text{Hf}_{15}$ for $\{110\}\langle1\bar{1}0\rangle$ and $\{110\}\langle1\bar{1}1\rangle$ oriented crack tips, respectively.  Consistent 
with the Rice theory, the crack propagates via cleavage bond breaking and remains sharp in NbMoTaW for both of these crack orientations.  For 
$\text{Nb}_{45}\text{Ta}_{25}\text{Ti}_{15}\text{Hf}_{15}$, the results in Fig. \ref{fig:110_110} are similar to those in Fig. \ref{fig:110_121}, with initial propagation accompanying 
the emission of dislocations (Fig. \ref{fig:110_110}(c)) and eventual blunting of the crack tip.  The behavior for $\text{Nb}_{45}\text{Ta}_{25}\text{Ti}_{15}\text{Hf}_{15}$ is somewhat 
more complex for $\{110\}\langle1\bar{1}1\rangle$ oriented crack tips, with fewer dislocation emissions observed with a lower degree of crack blunting and cleavage bond breaking observed in the later stages. 

\begin{figure*}
    \centering
    \includegraphics[width = 1.0\textwidth]{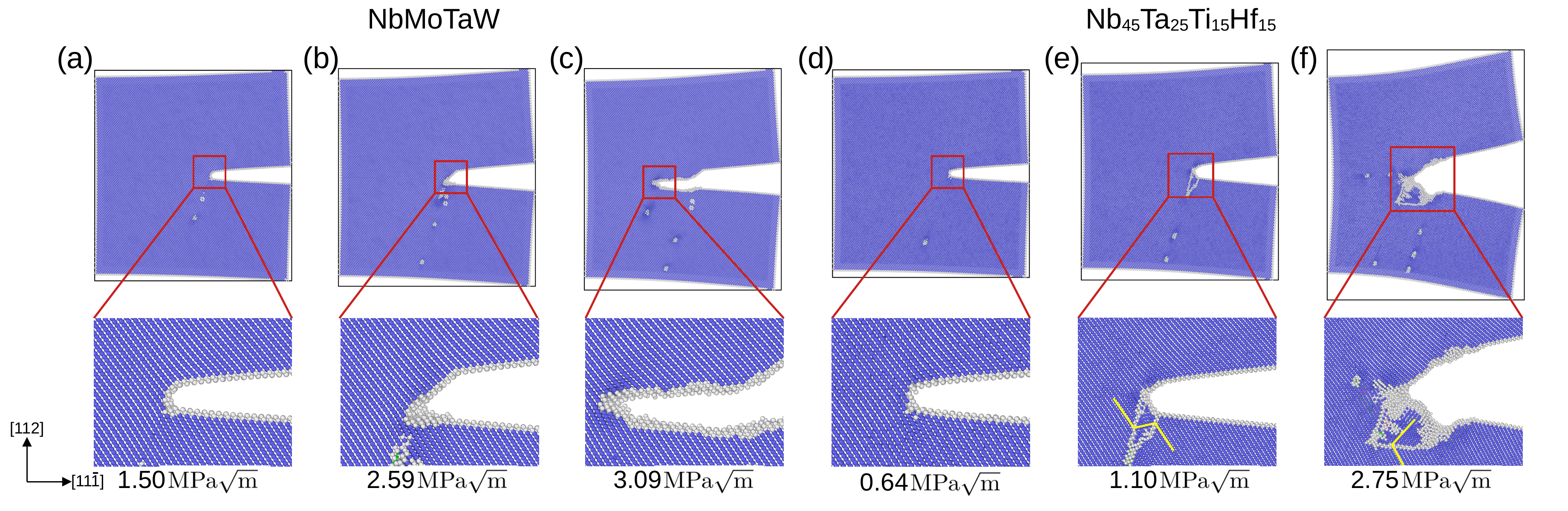}
    \caption{For a $\{112\}\langle1\bar{1}0\rangle$ crack orientation, atomic configurations from fracture simulations for NbMoTaW (a,b,c) and $\text{Nb}_{45}\text{Ta}_{25}\text{Ti}_{15}\text{Hf}_{15}$ (d,e,f), at different loadings represented by the values of $K_I$ listed below each panel.  Crack initiation events for an atomically sharp crack tip are provided in (a,d) and for later stages of crack propagation (b,c,e,f).  The crack initiates by dislocation emissions in NbMoTaW and $\text{Nb}_{45}\text{Ta}_{25}\text{Ti}_{15}\text{Hf}_{15}$. While the fracture mode in NbMoTaW switches to predominantly bond breaking, the fracture mode remains the same in $\text{Nb}_{45}\text{Ta}_{25}\text{Ti}_{15}\text{Hf}_{15}$ as crack propagation proceeds. The atoms are colored in the same manner as in Fig. \ref{fig:110_121}.}
    \label{fig:121_110}
\end{figure*}

We consider next the propagation of cracks normal to the $\{112\}$ surface, as shown in Fig. \ref{fig:121_110}. For this $\{112\}\langle1\bar{1}0\rangle$ crack orientation, MoNbTaW 
cracks no longer propagate in an atomically sharp manner, and display dislocation emission.  Similar to the initial stages of crack propagation in this system for 
the $\{110\}\langle1\bar{1}2\rangle$ crack orientation, this behavior for MoNbTaW is in contrast to the picture from the Rice theory, and may reflect the effects of compositional 
fluctuations arising from chemical complexity.  This dislocation emission does not lead to a high degree of crack blunting, however, and the crack propagation switches to predominantly 
bond breaking in later stages.  For $\text{Nb}_{45}\text{Ta}_{25}\text{Ti}_{15}\text{Hf}_{15}$ the behavior for $\{112\}$ fracture planes is similar to that observed for fracture on 
$\{110\}$ planes, with crack propagation accompanying dislocation emission, twin formation, and crack blunting.



\begin{figure*}
    \centering
    \includegraphics[width = 0.9\textwidth]{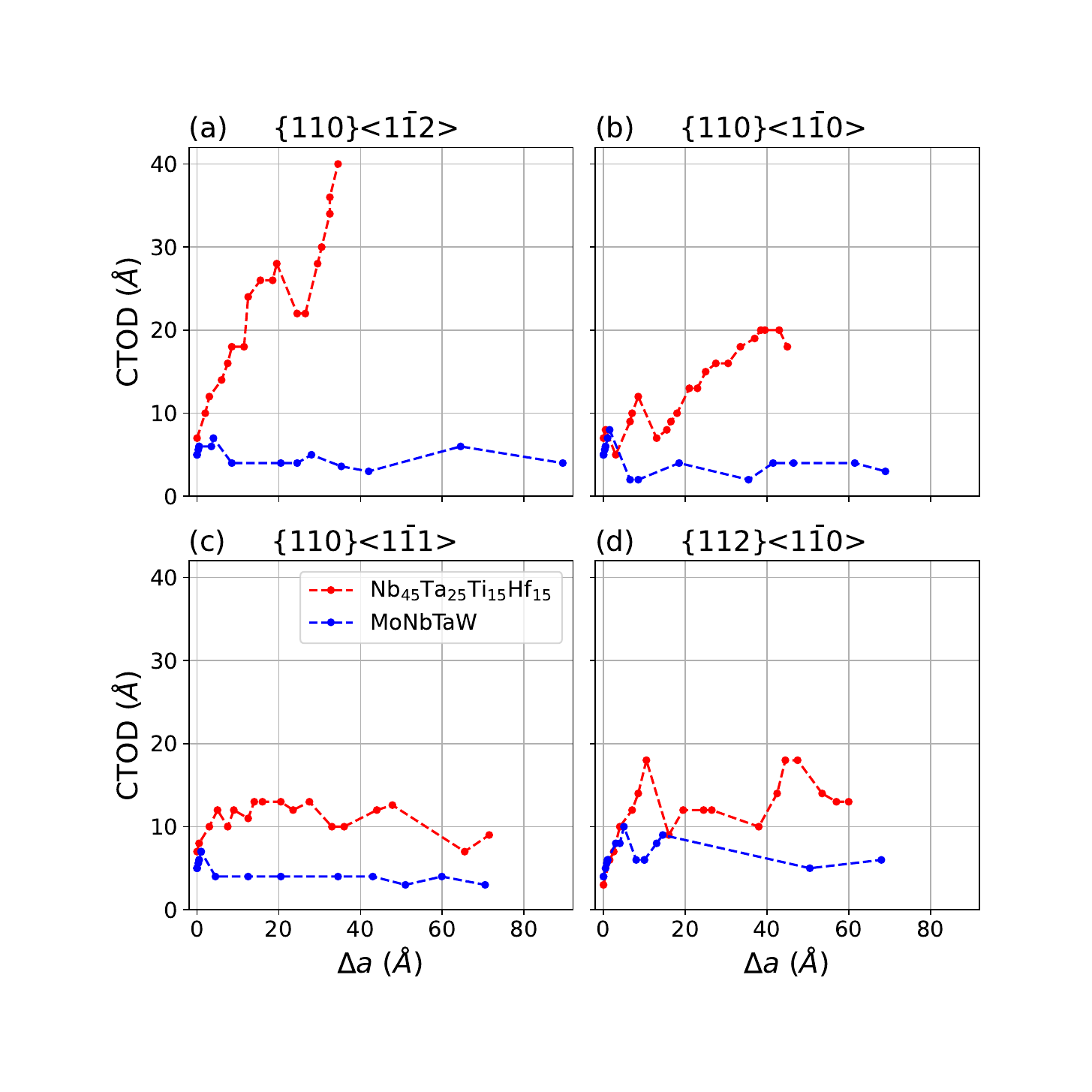}
    \caption{Calculated crack tip opening displacement (CTOD) as a function of crack extension ($\Delta$a) for $\text{Nb}_{45}\text{Ta}_{25}\text{Ti}_{15}\text{Hf}_{15}$ and NbMoTaW.  Results for each of the four different crack tip orientations are proviced in (a)-(d). The larger CTOD in $\text{Nb}_{45}\text{Ta}_{25}\text{Ti}_{15}\text{Hf}_{15}$ represents higher fracture resistance.}
    \label{fig:CTOD}
\end{figure*}

To further analyze the evolution of crack behavior once it begins to advance, we analyze the crack tip opening displacement (CTOD) as a function of the crack extension
$\Delta$a, reflecting a nanoscale version of the R-curve used in fracture mechanics.  In this analysis the CTOD is derived as distance between the opposite surfaces of the crack, the 
shape of which is approximated by that of a
hemisphere (see SI. Fig. S1-8 for details). Figure \ref{fig:CTOD} shows the atomistic crack-resistance R-curve, plotting CTOD versus $\Delta$a. The CTOD in 
$\text{Nb}_{45}\text{Ta}_{25}\text{Ti}_{15}\text{Hf}_{15}$ is higher than or approximately the same as that of NbMoTaW for all $\Delta$a values, for all crack orientations considered.
These results thus show good qualitative correspondence with the predictions of the Rice theory (Table I), suggesting more ductile behavior for the former RHEA. For all orientations 
considered, the crack tips in $\text{Nb}_{45}\text{Ta}_{25}\text{Ti}_{15}\text{Hf}_{15}$ involve a more irregular crack propagation path and a larger CTOD as the loading proceeds.
Those results indicate that a higher degree of crack tip blunting and fracture resistance for $\text{Nb}_{45}\text{Ta}_{25}\text{Ti}_{15}\text{Hf}_{15}$, also in qualitative agreement 
with the Rice theory.

\subsection{Analysis of Fracture Modes}

\begin{figure*}
    \centering
    \includegraphics[width = 0.9\textwidth]{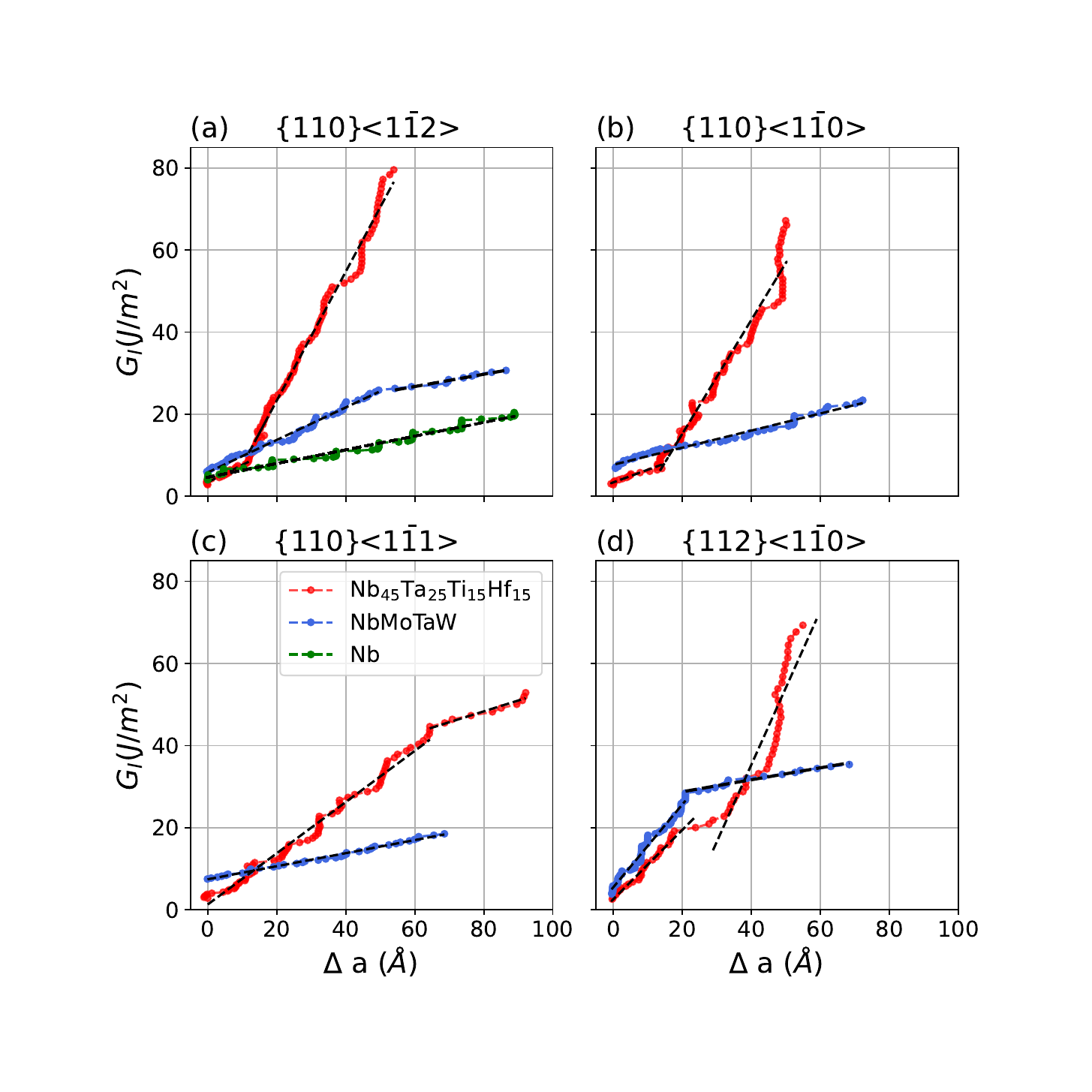}
    \caption{The calculated strain energy release ($G_I$) as a function of crack extension $\Delta$a in $\text{Nb}_{45}\text{Ta}_{25}\text{Ti}_{15}\text{Hf}_{15}$ and NbMoTaW with four different crack orientations, where higher slope corresponds to higher fracture resistance and the changes in slopes serves as an indicator for changes in the dominant fracture mode. These again are nanoscale R-curves.}
    \label{fig:G-R}
\end{figure*}

A notable finding from the results presented in the previous sub-section is that the mechanisms for crack-tip propagation (plasticity versus bond breaking) and geometry (blunted 
versus sharp) can differ between initial and later stages of crack propagation, starting from an atomically sharp crack tip. We consider next the origins of these changes, facilitated 
by an analysis of the strain energy release ($G_I$) as a function of crack tip extension ($\Delta$a).  A plot of $G_I$ versus $\Delta$a represents a nanoscale version of the 
$G_R$ curve considered in fracture mechanics, with higher slopes in this relationship corresponding to higher fracture resistance (i.e., more ductile behavior). To compute $G_I$ we make use of the following relationship with the stress intensity factor ($K_I$) under plane strain condition:
\begin{equation}
    G_I = K_I^2(1-\nu^2)/E
    \label{eq:gi}
\end{equation}
where $\nu$ is the Poisson's ratio and $E$ is the Young's modulus. As above, we determine values for the isotropic elastic moduli in Eq.~\ref{eq:gi} via the VRH averages derived from 
the calculated single-crystal elastic constants.  The resulting $G_R$ curves are plotted in Fig.~\ref{fig:G-R}.

We consider first the results for pure Nb, illustrated for a $\{110\}\langle1\bar{1}2\rangle$ crack orientation in Fig.~\ref{fig:G-R}(a).  The constant slope reflects the same operating
fracture mode, namely cleavage fracture, over the entire range of $\Delta$a considered.  We note that these results for pure Nb were obtained using the MLIP due to Yin et al. \cite{yin2021atomistic},
although a similar curve with a constant slope having a magnitude within 22 percent of that shown in Fig.~\ref{fig:G-R} is obtained for Nb using the MLIP developed in the
current work, see Supplemental Fig.S9. This result indicates that the differences between Nb and $\text{Nb}_{45}\text{Ta}_{25}\text{Ti}_{15}\text{Hf}_{15}$ in Fig.~\ref{fig:G-R} 
(discussed below) are not an artifact of using two separate MLIP models in the fracture simulations, but rather reflect the role of alloy chemistry.

For $\{110\}\langle1\bar{1}0\rangle$ and $\{110\}\langle1\bar{1}1\rangle$ crack orientations, the results for MoNbTaW in Fig.~\ref{fig:G-R} (b), (c) show behavior similar to pure Nb,
with a constant slope reflecting cleavage fracture for all values of $\Delta$a plotted. By contrast, results for MoNbTaW in Fig.~\ref{fig:G-R} (a), (d) for 
$\{110\}\langle1\bar{1}2\rangle$ and $\{112\}\langle1\bar{1}0\rangle$ orientations, as well as $\text{Nb}_{45}\text{Ta}_{25}\text{Ti}_{15}\text{Hf}_{15}$ for all orientations, show 
changes in slope reflecting different fracture modes for initial and later stages of crack propagation.  

Considering the results for $\text{Nb}_{45}\text{Ta}_{25}\text{Ti}_{15}\text{Hf}_{15}$, we find that although the crack tip propagates predominately by dislocation activity,
the changes in slopes in Fig.~\ref{fig:G-R} indicate different modes of plasticity.  From an analysis of the atomic configurations before and after the change in slopes, we 
identify two dislocation-assisted crack growth modes: nucleation-limited extension and slip-limited extension. In the nucleation-limited mode, a large change in the loading stress intensity factor ($\Delta K_I$) is required to 
nucleate a dislocation at the crack tip, but once it is formed, the dislocation propagates into the bulk relatively easily. In the slip-limited mode, abundant dislocation sources 
are active at the crack tip, and the nucleated dislocations interact with already emitted  dislocations, causing resistance to slip. It 
is noted that in this slip-limited mode, the emission of a dislocation into the bulk crystal is sometimes accompanied by the annihilation of others. The changes in the dominant 
mode of ductile fracture are captured by the sharp changes in the slope in the $G_R$ curves for $\text{Nb}_{45}\text{Ta}_{25}\text{Ti}_{15}\text{Hf}_{15}$.  With this insight, we detail 
below the changes in the fracture modes observed in the simulations.

For the $\{110\}\langle1\bar{1}2\rangle$ (Fig. \ref{fig:G-R}(a)) and $\{112\}\langle1\bar{1}0\rangle$ (Fig. \ref{fig:G-R}(d)) 
crack orientations, both RHEAs exhibit two distinct stages
of crack extension. In the initial stage for NbMoTaW, for both crack orientations, the crack extends by a mixture of dislocation nucleation and slip limited crack extension, while in 
the later stage a bond-breaking cleavage mode prevails. For $\text{Nb}_{45}\text{Ta}_{25}\text{Ti}_{15}\text{Hf}_{15}$, for both orientations, crack extension is initially dislocation-nucleation limited, and switches to slip-limited in later stages.  For the nucleation limited regime, the average changes in the loading stress intensity factor are $\Delta K_I \approx 0.13 \text{MPa}\sqrt{\text{m}}$ 
and $\Delta K_I \approx 0.12 \text{MPa}\sqrt{\text{m}}$ to nucleate a dislocation for $\{110\}\langle1\bar{1}2\rangle$ and $\{112\}\langle1\bar{1}0\rangle$, respectively;
subsequent dislocation emission occurs at intervals of $\Delta K_I \approx 0.05 \text{MPa}\sqrt{\text{m}}$ and $\Delta K_I \approx 0.04 \text{MPa}\sqrt{\text{m}}$, respectively.
The later crack extension stage for $\text{Nb}_{45}\text{Ta}_{25}\text{Ti}_{15}\text{Hf}_{15}$ is dislocation-slip limited, requiring on average
$\Delta K_I \approx 0.18 \text{MPa}\sqrt{\text{m}}$ and $\Delta K_I \approx 0.26 \text{MPa}\sqrt{\text{m}}$ for $\{110\}\langle1\bar{1}2\rangle$ and $\{112\}\langle1\bar{1}0\rangle$, respectively,
to advance an already existing dislocation, and one or more dislocations are nucleated almost immediately after such an event; in the interval between 
dislocation emissions, the number, total length, and the shape of the dislocations change as a result of their interaction. In the case of the $\{112\}\langle1\bar{1}0\rangle$ crack 
orientation, at a later stage a new grain is formed at the crack tip and the nearby grain boundary acts as a source for dislocations (see Fig.~\ref{fig:121_110}(f)).

Considering the $\{110\}\langle1\bar{1}0\rangle$ (Fig. \ref{fig:G-R}(b)), and $\{110\}\langle1\bar{1}1\rangle$ (Fig. \ref{fig:G-R}(c)) crack orientations, MoNbTaW displays a constant
slope corresponding to a cleavage fracture mode.  By contrast, $\text{Nb}_{45}\text{Ta}_{25}\text{Ti}_{15}\text{Hf}_{15}$ again displays two crack-extension stages.  For the 
$\{110\}\langle1\bar{1}0\rangle$ orientation, crack extension is initially dislocation-nucleation limited, and switches to slip-limited in later stages, similar to the orientations discussed above.  For the nucleation limited regime, the average loading is $\Delta K_I \approx 0.11 \text{MPa}\sqrt{\text{m}}$ to nucleate a dislocation.  Subsequent dislocation  
emission occurs at intervals of $\Delta K_I \approx 0.13 \text{MPa}\sqrt{\text{m}}$.
The later crack extension stage for $\text{Nb}_{45}\text{Ta}_{25}\text{Ti}_{15}\text{Hf}_{15}$ is dislocation-slip limited: on average, requiring 
$\Delta K_I \approx 0.17 \text{MPa}\sqrt{\text{m}}$ to advance an already existing dislocation.
For the $\{110\}\langle1\bar{1}1\rangle$ orientation, the initial stage (lower slope) corresponds to a slip-limited regime in which three dislocation emission events are observed. In
between these emission events, multiple dislocations interact, causing resistance to slip, and a higher slope in Fig.~\ref{fig:G-R}(c). At later stages, the crack-growth mode switches 
to cleavage (bond-breaking), with a correspondingly lower slope in Fig.~\ref{fig:G-R}(c).

\section{Discussion}

In summary, atomistic simulations based on machine-learning interatomic potentials to have been used to investigate differences in crack-propagation mechanisms in elemental Nb and two RHEAs:  $\text{Nb}_{45}\text{Ta}_{25}\text{Ti}_{15}\text{Hf}_{15}$ and NbMoTaW. Based on analyses of crack tip opening displacement relations and strain energy release rates derived from these simulations, we establish three crack 
extension modes: cleavage bond breaking, dislocation nucleation limited extension, and dislocation slip limited extension. For a given system and crack orientation, the fracture 
resistance is the highest (most ductile) in the dislocation slip limited crack extension mode, lowest (most brittle) in the cleavage bond breaking case, and intermediate for dislocation 
nucleation limited crack extension.

The results highlight important effects of high-entropy alloying on crack-tip behavior.  
Compared to elemental Nb, which displayed brittle cleavage fracture in the simulations, the RHEAs show a range of behavior, and the possibility of changes in fracture mode as the crack advances and changes shape.  In contrast to predictions of the Rice theory using averaged values of the unstable stacking fault energy, which indicate that MoNbTaW should be more
brittle than Nb, the former shows dislocation emission at the crack tip in some conditions, which we attribute to the presence of fluctuations in dislocation nucleation barriers induced
by the chemical complexity of the RHEA.  Consistent with the Rice theory predictions, $\text{Nb}_{45}\text{Ta}_{25}\text{Ti}_{15}\text{Hf}_{15}$ shows crack extension behavior consistent
with ductile fracture.  
For this system, alloying not only changes the fracture mode relative to elemental Nb, but promotes dislocation accumulation at the crack tip, leading to higher resistance to crack propagation in later stages as the crack tip becomes more blunted in shape.

The present work establishes important insights into the mechanisms by which alloying with multicomponent species can impact crack tip propagation in RHEAs.
They further suggest  important directions for future analysis.  Specifically, the present simulations are limited in their application of energy minimization calculations in systems with fixed crystal orientations.  
They thus represent low-temperature responses of single crystals.  
To advance understanding of fracture under more realistic conditions, extensions of these simulations to consider effects of grain boundaries and finite temperature are of interest, and can be pursued through molecular-dynamics simulations on polycrystalline models.  
Of additional interest is the role of deformation twinning which was observed to be present for some crack orientations.  
The growing availability of MLIPs for HEAs, including the one developed here for Nb-Ta-Ti-Hf, opens the opportunity for performing such simulations with accurate representations of the chemical interactions between multiple elements, and thus increasing the understanding of how chemical composition can be tuned in design of HEAs with targeted damage tolerance.

\section{Materials and Methods}
\subsection{Interatomic potential development}

\begin{figure*}
    \centering
    \includegraphics[width = 0.9\textwidth]{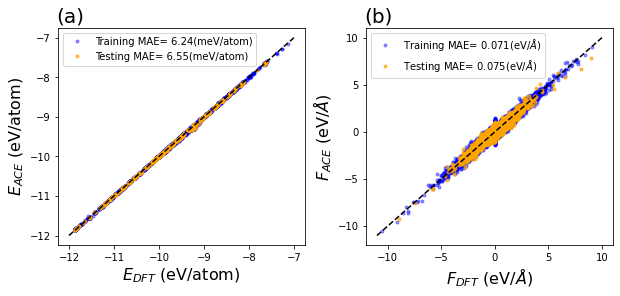}
    \caption{Parity plots comparing the ACE and DFT calculations for the energies (a) and forces (b), for both test and training data sets. The insets give the mean absolute error (MAE) for each value.}
    \label{fig:parity}
\end{figure*}

For the atomistic simulations we employed a previously published MLIP for NbMoTaW and Nb \cite{yin2021atomistic} based on the moment tensor potential (MTP) formalism 
\cite{zuo2020performance}. For NbTaTiHf, we have developed a new MLIP based on the atomic cluster expansion (ACE) formalism 
\cite{drautz2019atomic,lysogorskiy2021performant,bochkarev2022efficient,lysogorskiy2023active}. 
The ACE potential is trained on a large database of energies and forces derived from DFT calculations performed using the Vienna Ab-Initio Simulation Package (VASP) 
\cite{kresse1996efficient,kresse1993ab}. The initial dataset contains energies and forces for the configurations (beside those corresponding to the ab-initio molecular dynamics
simulations) used in the training of the MoNbTaW MLIP \cite{li2020complex}, as well as some configurations taken from the first-principles study of Borges et al. \cite{borges2024local}.
An initial ACE MLIP for NbTaTiHf was trained on these configurations and used to initiate an iterative refinement of the potential, employing an active learning process in which 
configurations with high values of the so-called extrapolation grade \cite{lysogorskiy2023active} were identified from finite-temperature molecular-dynamics (MD) simulations under periodic boundary conditions at elevated temperatures (1000K, 1600K, 2000K, and 2400K) and free surfaces at room temperature. 
These MD simulations and extrapolation-grade analysis were performed using the performant atomic cluster expansion (PACE) implementation \cite{lysogorskiy2021performant} in the Large-
scale Atomic/Molecular Massively Parallel Simulator (LAMMPS) software \cite{thompson2022lammps}. The configurations with high extrapolation grade were extracted for DFT calculations 
and added to the training set to refine the potential further. 
This iterative process continued until finite-temperature MD simulations produced no configurations with an extrapolation grade higher than 5.
For the final version of the potential, we considered a total of 6493 structures, and 263,810 atoms. 
The MLIP was trained on 90\% of the data, with the remaining used as a test set using the \textit{pacemaker} software \cite{lysogorskiy2021performant}.
The results of the training are summarized in Fig. \ref{fig:parity}, showing parity plots for energy (a) and forces (b), for both training and test sets.  Further details of the DFT
calculations and MLIP training are provided in Table SI-III in the Supplemental Information (SI), where additional results are presented for properties derived from the potential, including elastic 
constants, generalized stacking faults, local lattice distortion, etc.

\subsection{Fracture simulations}
The MLIPs were used in fracture simulations employing the $K$-test geometry with a semi-infinite crack \cite{andric2018atomistic}.  In these simulations, the response of a crack tip
is analyzed for incrementally increasing values of an applied stress intensity factor $K_I$, which sets the boundary conditions far away from the crack tip.
The simulation cells are approximately 55 $\AA$ thick along the crack line direction, 300 $\AA$ long in the direction of crack propagation, and 300 nm along the direction normal to the 
fracture plane.  The simulation cells contain from 278,000 to 314,000 atoms. Results highlighted in the main text are found to be relatively insensitive to further increases in 
simulation size.  As described in the text, four different orientations are considered for the fracture plane and crack line direction. These simulation geometries are chosen to align 
the crack surface along the bcc planes ($\{110\}$ and $\{112\}$). Periodic boundary conditions are used in the line direction of the crack tip with free surfaces employed normal to the 
other two directions. The crack is introduced by removing a layer of atoms to mimic the slightly blunted crack and to generate a traction free surface. The simulation cell is loaded via
the anisotropic elastic displacement field solution of linear elastic fracture mechanics (LEFM) \cite{sun2011fracture} under plane strain conditions, followed by the static relaxation 
of the atoms in the interior of the simulation cell; in these relaxations the atoms in the exterior (2$r_c$ from the boundary) are kept frozen 
at positions dictated by the LEFM solutions.
The simulations were repeated multiple times, incrementing the loading by $\Delta K = 0.01 \text{MPa}\sqrt{\text{m}}$ in each subsequent iteration. All static relaxations were performed
in LAMMPS using the ``FIRE" algorithm \cite{bitzek2006structural} with energy tolerance of $1\times10^{-9}$ and force tolerance of $1\times10^{-3}$ eV/$\AA$. 

\section{Acknowledgements}
\subsection{Funding and computational resources}
This work was supported by the U.S. Department of Energy, Office of Science, Office of Basic Energy Sciences, Materials Sciences and Engineering Division, under Contract No. DE-AC02-05-CH11231 within the Damage Tolerance in Structural Materials (KC 13) program.

The study made use of resources of the National Energy Research Scientific Computing Center (NERSC), a U.S. Department of Energy Office of Science User Facility located at Lawrence Berkeley National Laboratory, operated under the same contract number, using NERSC Award No. BES-ERCAP0027535.

This research used the Lawrencium computational cluster resource provided by the IT Division at the Lawrence Berkeley National Laboratory (Supported by the Director, Office of Science, Office of Basic Energy Sciences, of the U.S. Department of Energy under Contract No. DE-AC02-05CH11231)

The Scientific Computing Group (also known as High-Performance Computing Services) under Science IT supports the mission of Lawrence Berkeley National Laboratory by providing technology and consulting support for science and technical programs, in the areas of data management, HPC cluster computing, and Cloud services.

\bibliographystyle{apsrev4-2}
\bibliography{main.bbl.bib}

\begin{thebibliography}{38}%
\makeatletter
\providecommand \@ifxundefined [1]{%
 \@ifx{#1\undefined}
}%
\providecommand \@ifnum [1]{%
 \ifnum #1\expandafter \@firstoftwo
 \else \expandafter \@secondoftwo
 \fi
}%
\providecommand \@ifx [1]{%
 \ifx #1\expandafter \@firstoftwo
 \else \expandafter \@secondoftwo
 \fi
}%
\providecommand \natexlab [1]{#1}%
\providecommand \enquote  [1]{``#1''}%
\providecommand \bibnamefont  [1]{#1}%
\providecommand \bibfnamefont [1]{#1}%
\providecommand \citenamefont [1]{#1}%
\providecommand \href@noop [0]{\@secondoftwo}%
\providecommand \href [0]{\begingroup \@sanitize@url \@href}%
\providecommand \@href[1]{\@@startlink{#1}\@@href}%
\providecommand \@@href[1]{\endgroup#1\@@endlink}%
\providecommand \@sanitize@url [0]{\catcode `\\12\catcode `\$12\catcode `\&12\catcode `\#12\catcode `\^12\catcode `\_12\catcode `\%12\relax}%
\providecommand \@@startlink[1]{}%
\providecommand \@@endlink[0]{}%
\providecommand \url  [0]{\begingroup\@sanitize@url \@url }%
\providecommand \@url [1]{\endgroup\@href {#1}{\urlprefix }}%
\providecommand \urlprefix  [0]{URL }%
\providecommand \Eprint [0]{\href }%
\providecommand \doibase [0]{https://doi.org/}%
\providecommand \selectlanguage [0]{\@gobble}%
\providecommand \bibinfo  [0]{\@secondoftwo}%
\providecommand \bibfield  [0]{\@secondoftwo}%
\providecommand \translation [1]{[#1]}%
\providecommand \BibitemOpen [0]{}%
\providecommand \bibitemStop [0]{}%
\providecommand \bibitemNoStop [0]{.\EOS\space}%
\providecommand \EOS [0]{\spacefactor3000\relax}%
\providecommand \BibitemShut  [1]{\csname bibitem#1\endcsname}%
\let\auto@bib@innerbib\@empty
\bibitem [{\citenamefont {Senkov}\ \emph {et~al.}(2011)\citenamefont {Senkov}, \citenamefont {Wilks}, \citenamefont {Scott},\ and\ \citenamefont {Miracle}}]{senkov2011mechanical}%
  \BibitemOpen
  \bibfield  {author} {\bibinfo {author} {\bibfnamefont {O.~N.}\ \bibnamefont {Senkov}}, \bibinfo {author} {\bibfnamefont {G.}~\bibnamefont {Wilks}}, \bibinfo {author} {\bibfnamefont {J.}~\bibnamefont {Scott}},\ and\ \bibinfo {author} {\bibfnamefont {D.~B.}\ \bibnamefont {Miracle}},\ }\href@noop {} {\bibfield  {journal} {\bibinfo  {journal} {Intermetallics}\ }\textbf {\bibinfo {volume} {19}},\ \bibinfo {pages} {698} (\bibinfo {year} {2011})}\BibitemShut {NoStop}%
\bibitem [{\citenamefont {Senkov}\ \emph {et~al.}(2010)\citenamefont {Senkov}, \citenamefont {Wilks}, \citenamefont {Miracle}, \citenamefont {Chuang},\ and\ \citenamefont {Liaw}}]{SENKOV20101758}%
  \BibitemOpen
  \bibfield  {author} {\bibinfo {author} {\bibfnamefont {O.}~\bibnamefont {Senkov}}, \bibinfo {author} {\bibfnamefont {G.}~\bibnamefont {Wilks}}, \bibinfo {author} {\bibfnamefont {D.}~\bibnamefont {Miracle}}, \bibinfo {author} {\bibfnamefont {C.}~\bibnamefont {Chuang}},\ and\ \bibinfo {author} {\bibfnamefont {P.}~\bibnamefont {Liaw}},\ }\href {https://doi.org/https://doi.org/10.1016/j.intermet.2010.05.014} {\bibfield  {journal} {\bibinfo  {journal} {Intermetallics}\ }\textbf {\bibinfo {volume} {18}},\ \bibinfo {pages} {1758} (\bibinfo {year} {2010})}\BibitemShut {NoStop}%
\bibitem [{\citenamefont {Han}\ \emph {et~al.}(2018)\citenamefont {Han}, \citenamefont {Luan}, \citenamefont {Liu}, \citenamefont {Chen}, \citenamefont {Li}, \citenamefont {Shao},\ and\ \citenamefont {Yao}}]{han2018microstructures}%
  \BibitemOpen
  \bibfield  {author} {\bibinfo {author} {\bibfnamefont {Z.}~\bibnamefont {Han}}, \bibinfo {author} {\bibfnamefont {H.}~\bibnamefont {Luan}}, \bibinfo {author} {\bibfnamefont {X.}~\bibnamefont {Liu}}, \bibinfo {author} {\bibfnamefont {N.}~\bibnamefont {Chen}}, \bibinfo {author} {\bibfnamefont {X.}~\bibnamefont {Li}}, \bibinfo {author} {\bibfnamefont {Y.}~\bibnamefont {Shao}},\ and\ \bibinfo {author} {\bibfnamefont {K.}~\bibnamefont {Yao}},\ }\href@noop {} {\bibfield  {journal} {\bibinfo  {journal} {Materials Science and Engineering: A}\ }\textbf {\bibinfo {volume} {712}},\ \bibinfo {pages} {380} (\bibinfo {year} {2018})}\BibitemShut {NoStop}%
\bibitem [{\citenamefont {Kumar}\ \emph {et~al.}(2023)\citenamefont {Kumar}, \citenamefont {Kim}, \citenamefont {Yu}, \citenamefont {Ell}, \citenamefont {Zhang}, \citenamefont {Yang}, \citenamefont {Kim}, \citenamefont {Park}, \citenamefont {Minor}, \citenamefont {Park} \emph {et~al.}}]{kumar2023compressive}%
  \BibitemOpen
  \bibfield  {author} {\bibinfo {author} {\bibfnamefont {P.}~\bibnamefont {Kumar}}, \bibinfo {author} {\bibfnamefont {S.~J.}\ \bibnamefont {Kim}}, \bibinfo {author} {\bibfnamefont {Q.}~\bibnamefont {Yu}}, \bibinfo {author} {\bibfnamefont {J.}~\bibnamefont {Ell}}, \bibinfo {author} {\bibfnamefont {M.}~\bibnamefont {Zhang}}, \bibinfo {author} {\bibfnamefont {Y.}~\bibnamefont {Yang}}, \bibinfo {author} {\bibfnamefont {J.~Y.}\ \bibnamefont {Kim}}, \bibinfo {author} {\bibfnamefont {H.-K.}\ \bibnamefont {Park}}, \bibinfo {author} {\bibfnamefont {A.~M.}\ \bibnamefont {Minor}}, \bibinfo {author} {\bibfnamefont {E.~S.}\ \bibnamefont {Park}}, \emph {et~al.},\ }\href@noop {} {\bibfield  {journal} {\bibinfo  {journal} {Acta Materialia}\ }\textbf {\bibinfo {volume} {245}},\ \bibinfo {pages} {118620} (\bibinfo {year} {2023})}\BibitemShut {NoStop}%
\bibitem [{\citenamefont {Kumar}\ \emph {et~al.}(2024)\citenamefont {Kumar}, \citenamefont {Gou}, \citenamefont {Cook}, \citenamefont {Payne}, \citenamefont {Morrison}, \citenamefont {Wang}, \citenamefont {Zhang}, \citenamefont {Asta}, \citenamefont {Minor}, \citenamefont {Cao}, \citenamefont {Li},\ and\ \citenamefont {Ritchie}}]{Punit2024}%
  \BibitemOpen
  \bibfield  {author} {\bibinfo {author} {\bibfnamefont {P.}~\bibnamefont {Kumar}}, \bibinfo {author} {\bibfnamefont {X.}~\bibnamefont {Gou}}, \bibinfo {author} {\bibfnamefont {D.~H.}\ \bibnamefont {Cook}}, \bibinfo {author} {\bibfnamefont {M.~I.}\ \bibnamefont {Payne}}, \bibinfo {author} {\bibfnamefont {N.~J.}\ \bibnamefont {Morrison}}, \bibinfo {author} {\bibfnamefont {W.}~\bibnamefont {Wang}}, \bibinfo {author} {\bibfnamefont {M.}~\bibnamefont {Zhang}}, \bibinfo {author} {\bibfnamefont {M.}~\bibnamefont {Asta}}, \bibinfo {author} {\bibfnamefont {A.~M.}\ \bibnamefont {Minor}}, \bibinfo {author} {\bibfnamefont {R.}~\bibnamefont {Cao}}, \bibinfo {author} {\bibfnamefont {Y.}~\bibnamefont {Li}},\ and\ \bibinfo {author} {\bibfnamefont {R.~O.}\ \bibnamefont {Ritchie}},\ }\href@noop {} {\bibfield  {journal} {\bibinfo  {journal} {Acta Materialia}\ ,\ \bibinfo {pages} {120297}} (\bibinfo {year} {2024})}\BibitemShut {NoStop}%
\bibitem [{\citenamefont {Fan}\ \emph {et~al.}(2022)\citenamefont {Fan}, \citenamefont {Qu},\ and\ \citenamefont {Zhang}}]{fan2022remarkably}%
  \BibitemOpen
  \bibfield  {author} {\bibinfo {author} {\bibfnamefont {X.}~\bibnamefont {Fan}}, \bibinfo {author} {\bibfnamefont {R.}~\bibnamefont {Qu}},\ and\ \bibinfo {author} {\bibfnamefont {Z.}~\bibnamefont {Zhang}},\ }\href@noop {} {\bibfield  {journal} {\bibinfo  {journal} {Journal of Materials Science \& Technology}\ }\textbf {\bibinfo {volume} {123}},\ \bibinfo {pages} {70} (\bibinfo {year} {2022})}\BibitemShut {NoStop}%
\bibitem [{\citenamefont {Cook}\ \emph {et~al.}(2024)\citenamefont {Cook}, \citenamefont {Kumar}, \citenamefont {Payne}, \citenamefont {Belcher}, \citenamefont {Borges}, \citenamefont {Wang}, \citenamefont {Walsh}, \citenamefont {Li}, \citenamefont {Devaraj}, \citenamefont {Zhang} \emph {et~al.}}]{cook2024kink}%
  \BibitemOpen
  \bibfield  {author} {\bibinfo {author} {\bibfnamefont {D.~H.}\ \bibnamefont {Cook}}, \bibinfo {author} {\bibfnamefont {P.}~\bibnamefont {Kumar}}, \bibinfo {author} {\bibfnamefont {M.~I.}\ \bibnamefont {Payne}}, \bibinfo {author} {\bibfnamefont {C.~H.}\ \bibnamefont {Belcher}}, \bibinfo {author} {\bibfnamefont {P.}~\bibnamefont {Borges}}, \bibinfo {author} {\bibfnamefont {W.}~\bibnamefont {Wang}}, \bibinfo {author} {\bibfnamefont {F.}~\bibnamefont {Walsh}}, \bibinfo {author} {\bibfnamefont {Z.}~\bibnamefont {Li}}, \bibinfo {author} {\bibfnamefont {A.}~\bibnamefont {Devaraj}}, \bibinfo {author} {\bibfnamefont {M.}~\bibnamefont {Zhang}}, \emph {et~al.},\ }\href@noop {} {\bibfield  {journal} {\bibinfo  {journal} {Science}\ }\textbf {\bibinfo {volume} {384}},\ \bibinfo {pages} {178} (\bibinfo {year} {2024})}\BibitemShut {NoStop}%
\bibitem [{\citenamefont {Lin}\ and\ \citenamefont {Thomson}(1986)}]{lin1986cleavage}%
  \BibitemOpen
  \bibfield  {author} {\bibinfo {author} {\bibfnamefont {I.-H.}\ \bibnamefont {Lin}}\ and\ \bibinfo {author} {\bibfnamefont {R.}~\bibnamefont {Thomson}},\ }\href@noop {} {\bibfield  {journal} {\bibinfo  {journal} {Acta Metallurgica}\ }\textbf {\bibinfo {volume} {34}},\ \bibinfo {pages} {187} (\bibinfo {year} {1986})}\BibitemShut {NoStop}%
\bibitem [{\citenamefont {Rice}\ and\ \citenamefont {Thomson}(1974)}]{rice1974ductile}%
  \BibitemOpen
  \bibfield  {author} {\bibinfo {author} {\bibfnamefont {J.~R.}\ \bibnamefont {Rice}}\ and\ \bibinfo {author} {\bibfnamefont {R.}~\bibnamefont {Thomson}},\ }\href@noop {} {\bibfield  {journal} {\bibinfo  {journal} {The Philosophical Magazine: A Journal of Theoretical Experimental and Applied Physics}\ }\textbf {\bibinfo {volume} {29}},\ \bibinfo {pages} {73} (\bibinfo {year} {1974})}\BibitemShut {NoStop}%
\bibitem [{\citenamefont {Rice}(1992)}]{rice1992dislocation}%
  \BibitemOpen
  \bibfield  {author} {\bibinfo {author} {\bibfnamefont {J.~R.}\ \bibnamefont {Rice}},\ }\href@noop {} {\bibfield  {journal} {\bibinfo  {journal} {Journal of the Mechanics and Physics of Solids}\ }\textbf {\bibinfo {volume} {40}},\ \bibinfo {pages} {239} (\bibinfo {year} {1992})}\BibitemShut {NoStop}%
\bibitem [{\citenamefont {Borges}\ \emph {et~al.}(2024{\natexlab{a}})\citenamefont {Borges}, \citenamefont {Ritchie},\ and\ \citenamefont {Asta}}]{borges2024electronic}%
  \BibitemOpen
  \bibfield  {author} {\bibinfo {author} {\bibfnamefont {P.~P.}\ \bibnamefont {Borges}}, \bibinfo {author} {\bibfnamefont {R.~O.}\ \bibnamefont {Ritchie}},\ and\ \bibinfo {author} {\bibfnamefont {M.}~\bibnamefont {Asta}},\ }\href@noop {} {\bibfield  {journal} {\bibinfo  {journal} {Science Advances}\ }\textbf {\bibinfo {volume} {10}},\ \bibinfo {pages} {eadp7670} (\bibinfo {year} {2024}{\natexlab{a}})}\BibitemShut {NoStop}%
\bibitem [{\citenamefont {Tandoc}\ \emph {et~al.}()\citenamefont {Tandoc}, \citenamefont {Hu}, \citenamefont {Qi},\ and\ \citenamefont {Liaw}}]{tandocmining}%
  \BibitemOpen
  \bibfield  {author} {\bibinfo {author} {\bibfnamefont {C.}~\bibnamefont {Tandoc}}, \bibinfo {author} {\bibfnamefont {Y.}~\bibnamefont {Hu}}, \bibinfo {author} {\bibfnamefont {L.}~\bibnamefont {Qi}},\ and\ \bibinfo {author} {\bibfnamefont {P.}~\bibnamefont {Liaw}},\ }\href@noop {} {\bibinfo {title} {Mining of lattice distortion, strength, and intrinsic ductility of refractory high entropy alloys, npj. comput. mater. 9 (2023) 53}}\BibitemShut {NoStop}%
\bibitem [{\citenamefont {De~Jong}\ \emph {et~al.}(2017)\citenamefont {De~Jong}, \citenamefont {Winter}, \citenamefont {Chrzan},\ and\ \citenamefont {Asta}}]{de2017ideal}%
  \BibitemOpen
  \bibfield  {author} {\bibinfo {author} {\bibfnamefont {M.}~\bibnamefont {De~Jong}}, \bibinfo {author} {\bibfnamefont {I.}~\bibnamefont {Winter}}, \bibinfo {author} {\bibfnamefont {D.}~\bibnamefont {Chrzan}},\ and\ \bibinfo {author} {\bibfnamefont {M.}~\bibnamefont {Asta}},\ }\href@noop {} {\bibfield  {journal} {\bibinfo  {journal} {Physical Review B}\ }\textbf {\bibinfo {volume} {96}},\ \bibinfo {pages} {014105} (\bibinfo {year} {2017})}\BibitemShut {NoStop}%
\bibitem [{\citenamefont {Qi}\ and\ \citenamefont {Chrzan}(2014)}]{qi2014tuning}%
  \BibitemOpen
  \bibfield  {author} {\bibinfo {author} {\bibfnamefont {L.}~\bibnamefont {Qi}}\ and\ \bibinfo {author} {\bibfnamefont {D.}~\bibnamefont {Chrzan}},\ }\href@noop {} {\bibfield  {journal} {\bibinfo  {journal} {Physical review letters}\ }\textbf {\bibinfo {volume} {112}},\ \bibinfo {pages} {115503} (\bibinfo {year} {2014})}\BibitemShut {NoStop}%
\bibitem [{\citenamefont {Andric}\ and\ \citenamefont {Curtin}(2018)}]{andric2018atomistic}%
  \BibitemOpen
  \bibfield  {author} {\bibinfo {author} {\bibfnamefont {P.}~\bibnamefont {Andric}}\ and\ \bibinfo {author} {\bibfnamefont {W.~A.}\ \bibnamefont {Curtin}},\ }\href@noop {} {\bibfield  {journal} {\bibinfo  {journal} {Modelling and Simulation in Materials Science and Engineering}\ }\textbf {\bibinfo {volume} {27}},\ \bibinfo {pages} {013001} (\bibinfo {year} {2018})}\BibitemShut {NoStop}%
\bibitem [{\citenamefont {Li}\ \emph {et~al.}(2020{\natexlab{a}})\citenamefont {Li}, \citenamefont {Chen}, \citenamefont {He}, \citenamefont {Fang}, \citenamefont {Liu}, \citenamefont {Jiang}, \citenamefont {Liu}, \citenamefont {Yang},\ and\ \citenamefont {Liaw}}]{li2020unveiling}%
  \BibitemOpen
  \bibfield  {author} {\bibinfo {author} {\bibfnamefont {J.}~\bibnamefont {Li}}, \bibinfo {author} {\bibfnamefont {H.}~\bibnamefont {Chen}}, \bibinfo {author} {\bibfnamefont {Q.}~\bibnamefont {He}}, \bibinfo {author} {\bibfnamefont {Q.}~\bibnamefont {Fang}}, \bibinfo {author} {\bibfnamefont {B.}~\bibnamefont {Liu}}, \bibinfo {author} {\bibfnamefont {C.}~\bibnamefont {Jiang}}, \bibinfo {author} {\bibfnamefont {Y.}~\bibnamefont {Liu}}, \bibinfo {author} {\bibfnamefont {Y.}~\bibnamefont {Yang}},\ and\ \bibinfo {author} {\bibfnamefont {P.~K.}\ \bibnamefont {Liaw}},\ }\href@noop {} {\bibfield  {journal} {\bibinfo  {journal} {Physical Review Materials}\ }\textbf {\bibinfo {volume} {4}},\ \bibinfo {pages} {103612} (\bibinfo {year} {2020}{\natexlab{a}})}\BibitemShut {NoStop}%
\bibitem [{\citenamefont {Farkas}(2024)}]{farkas2024role}%
  \BibitemOpen
  \bibfield  {author} {\bibinfo {author} {\bibfnamefont {D.}~\bibnamefont {Farkas}},\ }\href@noop {} {\bibfield  {journal} {\bibinfo  {journal} {Computational Materials Science}\ }\textbf {\bibinfo {volume} {235}},\ \bibinfo {pages} {112758} (\bibinfo {year} {2024})}\BibitemShut {NoStop}%
\bibitem [{\citenamefont {Cheung}\ and\ \citenamefont {Yip}(1994)}]{cheung1994molecular}%
  \BibitemOpen
  \bibfield  {author} {\bibinfo {author} {\bibfnamefont {K.~S.}\ \bibnamefont {Cheung}}\ and\ \bibinfo {author} {\bibfnamefont {S.}~\bibnamefont {Yip}},\ }\href@noop {} {\bibfield  {journal} {\bibinfo  {journal} {Modelling and Simulation in Materials Science and Engineering}\ }\textbf {\bibinfo {volume} {2}},\ \bibinfo {pages} {865} (\bibinfo {year} {1994})}\BibitemShut {NoStop}%
\bibitem [{\citenamefont {Farkas}(1998)}]{farkas1998atomistic}%
  \BibitemOpen
  \bibfield  {author} {\bibinfo {author} {\bibfnamefont {D.}~\bibnamefont {Farkas}},\ }\href@noop {} {\bibfield  {journal} {\bibinfo  {journal} {Materials Science and Engineering: A}\ }\textbf {\bibinfo {volume} {249}},\ \bibinfo {pages} {249} (\bibinfo {year} {1998})}\BibitemShut {NoStop}%
\bibitem [{\citenamefont {Farkas}(2005)}]{farkas2005twinning}%
  \BibitemOpen
  \bibfield  {author} {\bibinfo {author} {\bibfnamefont {D.}~\bibnamefont {Farkas}},\ }\href@noop {} {\bibfield  {journal} {\bibinfo  {journal} {Philosophical Magazine}\ }\textbf {\bibinfo {volume} {85}},\ \bibinfo {pages} {387} (\bibinfo {year} {2005})}\BibitemShut {NoStop}%
\bibitem [{\citenamefont {Guo}\ and\ \citenamefont {Zhao}(2007)}]{guo2007atomistic}%
  \BibitemOpen
  \bibfield  {author} {\bibinfo {author} {\bibfnamefont {Y.-F.}\ \bibnamefont {Guo}}\ and\ \bibinfo {author} {\bibfnamefont {D.-L.}\ \bibnamefont {Zhao}},\ }\href@noop {} {\bibfield  {journal} {\bibinfo  {journal} {Materials Science and Engineering: A}\ }\textbf {\bibinfo {volume} {448}},\ \bibinfo {pages} {281} (\bibinfo {year} {2007})}\BibitemShut {NoStop}%
\bibitem [{\citenamefont {Yin}\ \emph {et~al.}(2021)\citenamefont {Yin}, \citenamefont {Zuo}, \citenamefont {Abu-Odeh}, \citenamefont {Zheng}, \citenamefont {Li}, \citenamefont {Ding}, \citenamefont {Ong}, \citenamefont {Asta},\ and\ \citenamefont {Ritchie}}]{yin2021atomistic}%
  \BibitemOpen
  \bibfield  {author} {\bibinfo {author} {\bibfnamefont {S.}~\bibnamefont {Yin}}, \bibinfo {author} {\bibfnamefont {Y.}~\bibnamefont {Zuo}}, \bibinfo {author} {\bibfnamefont {A.}~\bibnamefont {Abu-Odeh}}, \bibinfo {author} {\bibfnamefont {H.}~\bibnamefont {Zheng}}, \bibinfo {author} {\bibfnamefont {X.-G.}\ \bibnamefont {Li}}, \bibinfo {author} {\bibfnamefont {J.}~\bibnamefont {Ding}}, \bibinfo {author} {\bibfnamefont {S.~P.}\ \bibnamefont {Ong}}, \bibinfo {author} {\bibfnamefont {M.}~\bibnamefont {Asta}},\ and\ \bibinfo {author} {\bibfnamefont {R.~O.}\ \bibnamefont {Ritchie}},\ }\href@noop {} {\bibfield  {journal} {\bibinfo  {journal} {Nature communications}\ }\textbf {\bibinfo {volume} {12}},\ \bibinfo {pages} {4873} (\bibinfo {year} {2021})}\BibitemShut {NoStop}%
\bibitem [{\citenamefont {Rao}\ \emph {et~al.}(2019)\citenamefont {Rao}, \citenamefont {Akdim}, \citenamefont {Antillon}, \citenamefont {Woodward}, \citenamefont {Parthasarathy},\ and\ \citenamefont {Senkov}}]{rao2019modeling}%
  \BibitemOpen
  \bibfield  {author} {\bibinfo {author} {\bibfnamefont {S.}~\bibnamefont {Rao}}, \bibinfo {author} {\bibfnamefont {B.}~\bibnamefont {Akdim}}, \bibinfo {author} {\bibfnamefont {E.}~\bibnamefont {Antillon}}, \bibinfo {author} {\bibfnamefont {C.}~\bibnamefont {Woodward}}, \bibinfo {author} {\bibfnamefont {T.}~\bibnamefont {Parthasarathy}},\ and\ \bibinfo {author} {\bibfnamefont {O.~N.}\ \bibnamefont {Senkov}},\ }\href@noop {} {\bibfield  {journal} {\bibinfo  {journal} {Acta Materialia}\ }\textbf {\bibinfo {volume} {168}},\ \bibinfo {pages} {222} (\bibinfo {year} {2019})}\BibitemShut {NoStop}%
\bibitem [{\citenamefont {Tsuru}\ \emph {et~al.}(2024)\citenamefont {Tsuru}, \citenamefont {Han}, \citenamefont {Matsuura}, \citenamefont {Chen}, \citenamefont {Kishida}, \citenamefont {Iobzenko}, \citenamefont {Rao}, \citenamefont {Woodward}, \citenamefont {George},\ and\ \citenamefont {Inui}}]{tsuru2024intrinsic}%
  \BibitemOpen
  \bibfield  {author} {\bibinfo {author} {\bibfnamefont {T.}~\bibnamefont {Tsuru}}, \bibinfo {author} {\bibfnamefont {S.}~\bibnamefont {Han}}, \bibinfo {author} {\bibfnamefont {S.}~\bibnamefont {Matsuura}}, \bibinfo {author} {\bibfnamefont {Z.}~\bibnamefont {Chen}}, \bibinfo {author} {\bibfnamefont {K.}~\bibnamefont {Kishida}}, \bibinfo {author} {\bibfnamefont {I.}~\bibnamefont {Iobzenko}}, \bibinfo {author} {\bibfnamefont {S.~I.}\ \bibnamefont {Rao}}, \bibinfo {author} {\bibfnamefont {C.}~\bibnamefont {Woodward}}, \bibinfo {author} {\bibfnamefont {E.~P.}\ \bibnamefont {George}},\ and\ \bibinfo {author} {\bibfnamefont {H.}~\bibnamefont {Inui}},\ }\href@noop {} {\bibfield  {journal} {\bibinfo  {journal} {Nature Communications}\ }\textbf {\bibinfo {volume} {15}},\ \bibinfo {pages} {1706} (\bibinfo {year} {2024})}\BibitemShut {NoStop}%
\bibitem [{\citenamefont {Stukowski}\ \emph {et~al.}(2012)\citenamefont {Stukowski}, \citenamefont {Bulatov},\ and\ \citenamefont {Arsenlis}}]{stukowski2012automated}%
  \BibitemOpen
  \bibfield  {author} {\bibinfo {author} {\bibfnamefont {A.}~\bibnamefont {Stukowski}}, \bibinfo {author} {\bibfnamefont {V.~V.}\ \bibnamefont {Bulatov}},\ and\ \bibinfo {author} {\bibfnamefont {A.}~\bibnamefont {Arsenlis}},\ }\href@noop {} {\bibfield  {journal} {\bibinfo  {journal} {Modelling and Simulation in Materials Science and Engineering}\ }\textbf {\bibinfo {volume} {20}},\ \bibinfo {pages} {085007} (\bibinfo {year} {2012})}\BibitemShut {NoStop}%
\bibitem [{\citenamefont {Stukowski}(2009)}]{stukowski2009visualization}%
  \BibitemOpen
  \bibfield  {author} {\bibinfo {author} {\bibfnamefont {A.}~\bibnamefont {Stukowski}},\ }\href@noop {} {\bibfield  {journal} {\bibinfo  {journal} {Modelling and simulation in materials science and engineering}\ }\textbf {\bibinfo {volume} {18}},\ \bibinfo {pages} {015012} (\bibinfo {year} {2009})}\BibitemShut {NoStop}%
\bibitem [{\citenamefont {Zuo}\ \emph {et~al.}(2020)\citenamefont {Zuo}, \citenamefont {Chen}, \citenamefont {Li}, \citenamefont {Deng}, \citenamefont {Chen}, \citenamefont {Behler}, \citenamefont {Cs{\'a}nyi}, \citenamefont {Shapeev}, \citenamefont {Thompson}, \citenamefont {Wood} \emph {et~al.}}]{zuo2020performance}%
  \BibitemOpen
  \bibfield  {author} {\bibinfo {author} {\bibfnamefont {Y.}~\bibnamefont {Zuo}}, \bibinfo {author} {\bibfnamefont {C.}~\bibnamefont {Chen}}, \bibinfo {author} {\bibfnamefont {X.}~\bibnamefont {Li}}, \bibinfo {author} {\bibfnamefont {Z.}~\bibnamefont {Deng}}, \bibinfo {author} {\bibfnamefont {Y.}~\bibnamefont {Chen}}, \bibinfo {author} {\bibfnamefont {J.}~\bibnamefont {Behler}}, \bibinfo {author} {\bibfnamefont {G.}~\bibnamefont {Cs{\'a}nyi}}, \bibinfo {author} {\bibfnamefont {A.~V.}\ \bibnamefont {Shapeev}}, \bibinfo {author} {\bibfnamefont {A.~P.}\ \bibnamefont {Thompson}}, \bibinfo {author} {\bibfnamefont {M.~A.}\ \bibnamefont {Wood}}, \emph {et~al.},\ }\href@noop {} {\bibfield  {journal} {\bibinfo  {journal} {The Journal of Physical Chemistry A}\ }\textbf {\bibinfo {volume} {124}},\ \bibinfo {pages} {731} (\bibinfo {year} {2020})}\BibitemShut {NoStop}%
\bibitem [{\citenamefont {Drautz}(2019)}]{drautz2019atomic}%
  \BibitemOpen
  \bibfield  {author} {\bibinfo {author} {\bibfnamefont {R.}~\bibnamefont {Drautz}},\ }\href@noop {} {\bibfield  {journal} {\bibinfo  {journal} {Physical Review B}\ }\textbf {\bibinfo {volume} {99}},\ \bibinfo {pages} {014104} (\bibinfo {year} {2019})}\BibitemShut {NoStop}%
\bibitem [{\citenamefont {Lysogorskiy}\ \emph {et~al.}(2021)\citenamefont {Lysogorskiy}, \citenamefont {Oord}, \citenamefont {Bochkarev}, \citenamefont {Menon}, \citenamefont {Rinaldi}, \citenamefont {Hammerschmidt}, \citenamefont {Mrovec}, \citenamefont {Thompson}, \citenamefont {Cs{\'a}nyi}, \citenamefont {Ortner} \emph {et~al.}}]{lysogorskiy2021performant}%
  \BibitemOpen
  \bibfield  {author} {\bibinfo {author} {\bibfnamefont {Y.}~\bibnamefont {Lysogorskiy}}, \bibinfo {author} {\bibfnamefont {C.~v.~d.}\ \bibnamefont {Oord}}, \bibinfo {author} {\bibfnamefont {A.}~\bibnamefont {Bochkarev}}, \bibinfo {author} {\bibfnamefont {S.}~\bibnamefont {Menon}}, \bibinfo {author} {\bibfnamefont {M.}~\bibnamefont {Rinaldi}}, \bibinfo {author} {\bibfnamefont {T.}~\bibnamefont {Hammerschmidt}}, \bibinfo {author} {\bibfnamefont {M.}~\bibnamefont {Mrovec}}, \bibinfo {author} {\bibfnamefont {A.}~\bibnamefont {Thompson}}, \bibinfo {author} {\bibfnamefont {G.}~\bibnamefont {Cs{\'a}nyi}}, \bibinfo {author} {\bibfnamefont {C.}~\bibnamefont {Ortner}}, \emph {et~al.},\ }\href@noop {} {\bibfield  {journal} {\bibinfo  {journal} {npj computational materials}\ }\textbf {\bibinfo {volume} {7}},\ \bibinfo {pages} {97} (\bibinfo {year} {2021})}\BibitemShut {NoStop}%
\bibitem [{\citenamefont {Bochkarev}\ \emph {et~al.}(2022)\citenamefont {Bochkarev}, \citenamefont {Lysogorskiy}, \citenamefont {Menon}, \citenamefont {Qamar}, \citenamefont {Mrovec},\ and\ \citenamefont {Drautz}}]{bochkarev2022efficient}%
  \BibitemOpen
  \bibfield  {author} {\bibinfo {author} {\bibfnamefont {A.}~\bibnamefont {Bochkarev}}, \bibinfo {author} {\bibfnamefont {Y.}~\bibnamefont {Lysogorskiy}}, \bibinfo {author} {\bibfnamefont {S.}~\bibnamefont {Menon}}, \bibinfo {author} {\bibfnamefont {M.}~\bibnamefont {Qamar}}, \bibinfo {author} {\bibfnamefont {M.}~\bibnamefont {Mrovec}},\ and\ \bibinfo {author} {\bibfnamefont {R.}~\bibnamefont {Drautz}},\ }\href@noop {} {\bibfield  {journal} {\bibinfo  {journal} {Physical Review Materials}\ }\textbf {\bibinfo {volume} {6}},\ \bibinfo {pages} {013804} (\bibinfo {year} {2022})}\BibitemShut {NoStop}%
\bibitem [{\citenamefont {Lysogorskiy}\ \emph {et~al.}(2023)\citenamefont {Lysogorskiy}, \citenamefont {Bochkarev}, \citenamefont {Mrovec},\ and\ \citenamefont {Drautz}}]{lysogorskiy2023active}%
  \BibitemOpen
  \bibfield  {author} {\bibinfo {author} {\bibfnamefont {Y.}~\bibnamefont {Lysogorskiy}}, \bibinfo {author} {\bibfnamefont {A.}~\bibnamefont {Bochkarev}}, \bibinfo {author} {\bibfnamefont {M.}~\bibnamefont {Mrovec}},\ and\ \bibinfo {author} {\bibfnamefont {R.}~\bibnamefont {Drautz}},\ }\href@noop {} {\bibfield  {journal} {\bibinfo  {journal} {Physical Review Materials}\ }\textbf {\bibinfo {volume} {7}},\ \bibinfo {pages} {043801} (\bibinfo {year} {2023})}\BibitemShut {NoStop}%
\bibitem [{\citenamefont {Kresse}\ and\ \citenamefont {Furthm{\"u}ller}(1996)}]{kresse1996efficient}%
  \BibitemOpen
  \bibfield  {author} {\bibinfo {author} {\bibfnamefont {G.}~\bibnamefont {Kresse}}\ and\ \bibinfo {author} {\bibfnamefont {J.}~\bibnamefont {Furthm{\"u}ller}},\ }\href@noop {} {\bibfield  {journal} {\bibinfo  {journal} {Physical review B}\ }\textbf {\bibinfo {volume} {54}},\ \bibinfo {pages} {11169} (\bibinfo {year} {1996})}\BibitemShut {NoStop}%
\bibitem [{\citenamefont {Kresse}\ and\ \citenamefont {Hafner}(1993)}]{kresse1993ab}%
  \BibitemOpen
  \bibfield  {author} {\bibinfo {author} {\bibfnamefont {G.}~\bibnamefont {Kresse}}\ and\ \bibinfo {author} {\bibfnamefont {J.}~\bibnamefont {Hafner}},\ }\href@noop {} {\bibfield  {journal} {\bibinfo  {journal} {Physical review B}\ }\textbf {\bibinfo {volume} {47}},\ \bibinfo {pages} {558} (\bibinfo {year} {1993})}\BibitemShut {NoStop}%
\bibitem [{\citenamefont {Li}\ \emph {et~al.}(2020{\natexlab{b}})\citenamefont {Li}, \citenamefont {Chen}, \citenamefont {Zheng}, \citenamefont {Zuo},\ and\ \citenamefont {Ong}}]{li2020complex}%
  \BibitemOpen
  \bibfield  {author} {\bibinfo {author} {\bibfnamefont {X.-G.}\ \bibnamefont {Li}}, \bibinfo {author} {\bibfnamefont {C.}~\bibnamefont {Chen}}, \bibinfo {author} {\bibfnamefont {H.}~\bibnamefont {Zheng}}, \bibinfo {author} {\bibfnamefont {Y.}~\bibnamefont {Zuo}},\ and\ \bibinfo {author} {\bibfnamefont {S.~P.}\ \bibnamefont {Ong}},\ }\href@noop {} {\bibfield  {journal} {\bibinfo  {journal} {npj Computational Materials}\ }\textbf {\bibinfo {volume} {6}},\ \bibinfo {pages} {70} (\bibinfo {year} {2020}{\natexlab{b}})}\BibitemShut {NoStop}%
\bibitem [{\citenamefont {Borges}\ \emph {et~al.}(2024{\natexlab{b}})\citenamefont {Borges}, \citenamefont {Ritchie},\ and\ \citenamefont {Asta}}]{borges2024local}%
  \BibitemOpen
  \bibfield  {author} {\bibinfo {author} {\bibfnamefont {P.~P.}\ \bibnamefont {Borges}}, \bibinfo {author} {\bibfnamefont {R.~O.}\ \bibnamefont {Ritchie}},\ and\ \bibinfo {author} {\bibfnamefont {M.}~\bibnamefont {Asta}},\ }\href@noop {} {\bibfield  {journal} {\bibinfo  {journal} {Acta Materialia}\ }\textbf {\bibinfo {volume} {262}},\ \bibinfo {pages} {119415} (\bibinfo {year} {2024}{\natexlab{b}})}\BibitemShut {NoStop}%
\bibitem [{\citenamefont {Thompson}\ \emph {et~al.}(2022)\citenamefont {Thompson}, \citenamefont {Aktulga}, \citenamefont {Berger}, \citenamefont {Bolintineanu}, \citenamefont {Brown}, \citenamefont {Crozier}, \citenamefont {in't Veld}, \citenamefont {Kohlmeyer}, \citenamefont {Moore}, \citenamefont {Nguyen} \emph {et~al.}}]{thompson2022lammps}%
  \BibitemOpen
  \bibfield  {author} {\bibinfo {author} {\bibfnamefont {A.~P.}\ \bibnamefont {Thompson}}, \bibinfo {author} {\bibfnamefont {H.~M.}\ \bibnamefont {Aktulga}}, \bibinfo {author} {\bibfnamefont {R.}~\bibnamefont {Berger}}, \bibinfo {author} {\bibfnamefont {D.~S.}\ \bibnamefont {Bolintineanu}}, \bibinfo {author} {\bibfnamefont {W.~M.}\ \bibnamefont {Brown}}, \bibinfo {author} {\bibfnamefont {P.~S.}\ \bibnamefont {Crozier}}, \bibinfo {author} {\bibfnamefont {P.~J.}\ \bibnamefont {in't Veld}}, \bibinfo {author} {\bibfnamefont {A.}~\bibnamefont {Kohlmeyer}}, \bibinfo {author} {\bibfnamefont {S.~G.}\ \bibnamefont {Moore}}, \bibinfo {author} {\bibfnamefont {T.~D.}\ \bibnamefont {Nguyen}}, \emph {et~al.},\ }\href@noop {} {\bibfield  {journal} {\bibinfo  {journal} {Computer Physics Communications}\ }\textbf {\bibinfo {volume} {271}},\ \bibinfo {pages} {108171} (\bibinfo {year} {2022})}\BibitemShut {NoStop}%
\bibitem [{\citenamefont {Sun}\ and\ \citenamefont {Jin}(2011)}]{sun2011fracture}%
  \BibitemOpen
  \bibfield  {author} {\bibinfo {author} {\bibfnamefont {C.-T.}\ \bibnamefont {Sun}}\ and\ \bibinfo {author} {\bibfnamefont {Z.}~\bibnamefont {Jin}},\ }\href@noop {} {\emph {\bibinfo {title} {Fracture mechanics}}}\ (\bibinfo  {publisher} {Academic press},\ \bibinfo {year} {2011})\BibitemShut {NoStop}%
\bibitem [{\citenamefont {Bitzek}\ \emph {et~al.}(2006)\citenamefont {Bitzek}, \citenamefont {Koskinen}, \citenamefont {G{\"a}hler}, \citenamefont {Moseler},\ and\ \citenamefont {Gumbsch}}]{bitzek2006structural}%
  \BibitemOpen
  \bibfield  {author} {\bibinfo {author} {\bibfnamefont {E.}~\bibnamefont {Bitzek}}, \bibinfo {author} {\bibfnamefont {P.}~\bibnamefont {Koskinen}}, \bibinfo {author} {\bibfnamefont {F.}~\bibnamefont {G{\"a}hler}}, \bibinfo {author} {\bibfnamefont {M.}~\bibnamefont {Moseler}},\ and\ \bibinfo {author} {\bibfnamefont {P.}~\bibnamefont {Gumbsch}},\ }\href@noop {} {\bibfield  {journal} {\bibinfo  {journal} {Physical review letters}\ }\textbf {\bibinfo {volume} {97}},\ \bibinfo {pages} {170201} (\bibinfo {year} {2006})}\BibitemShut {NoStop}%
\end{thebibliography}%
\end{document}